\documentclass[%
 reprint,
 superscriptaddress,
 amsmath,amssymb,
 aps,
]{revtex4-1}
\usepackage[colorlinks, citecolor=red]{hyperref}
\usepackage{graphicx}% Include figure files
\usepackage{dcolumn}% Align table columns on decimal point
\usepackage{bm}% bold math
\usepackage{listings}
\usepackage{color}
\usepackage{subfigure}

\definecolor{dkgreen}{rgb}{0,0.6,0}
\definecolor{gray}{rgb}{0.5,0.5,0.5}
\definecolor{mauve}{rgb}{0.58,0,0.82}

\begin{document}

%\preprint{APS/123-QED}

\title{VQNet: Library for a Quantum-Classical Hybrid Neural Network}% Force line breaks with \\
%\thanks{}
\author{Zhao-Yun Chen}
 \email{czy@originqc.com}%Lines break automatically or can be forced with \\
 \affiliation{Key Laboratory of Quantum Information, CAS}
 \affiliation{Origin Quantum Computing, Hefei, China}
 
\author{Cheng Xue}
 \email{itachi@mail.ustc.edu.cn}
 \affiliation{Key Laboratory of Quantum Information, CAS}
\author{Si-Ming Chen}
 \affiliation{Key Laboratory of Quantum Information, CAS}
\author{Guo-Ping Guo}
 \email{gpguo@ustc.edu.cn} 
 \affiliation{Key Laboratory of Quantum Information, CAS}

\begin{abstract}

    Deep learning is a modern approach to realize artificial intelligence. Many frameworks exist to implement the machine learning task; however, performance is limited by computing resources.  Using a quantum computer to accelerate training is a promising approach. The variational quantum circuit (VQC) has gained a great deal of attention because it can be run on near-term quantum computers. In this paper, we establish a new framework that merges traditional machine learning tasks with the VQC. Users can implement a trainable quantum operation into a neural network. This framework enables the training of a quantum-classical hybrid task and may lead to a new area of quantum machine learning.
    
\iffalse
\begin{description}
\item[Usage]
Secondary publications and information retrieval purposes.
\item[PACS numbers]
May be entered using the \verb+\pacs{#1}+ command.
\item[Structure]
You may use the \texttt{description} environment to structure your abstract;
use the optional argument of the \verb+\item+ command to give the category of each item. 
\end{description}
\fi

\end{abstract}

\maketitle

%\tableofcontents

\section{Introduction}

Quantum computing provides an approach to revolutionize the computer. A quantum computer uses the features of quantum theory -- superposition and entanglement -- to perform completely different computing from classical computers\cite{feynman_simulating_1982}. Several works have proved that the quantum computer completely outperforms the classical computer for some problems, which is referred to as “quantum supremacy” or “quantum advantage”\cite{quantum_supremacy}. Although the principle of the quantum computer is clear and some prototype algorithms, such as Grover's search algorithm and Shor's factoring algorithm, have been proposed\cite{grover_quantum_1997, shor_polynomial-time_1999}, the realization of the quantum computer has many engineering problems. At present, chips with a maximum of approximately 70 qubits are made; however, all the qubits are too noisy to perform even one fault-tolerant operation. It is believed that the perfect fault-tolerant quantum computers will not be available in the foreseeable future.

Instead, many researches believe that an imperfect quantum computer can be realized in the near future, which is referred to as a noisy intermediate-scale quantum (NISQ) device, with hundreds of qubits and an error rate of approximately 0.1\%\cite{preskill_quantum_2018}. The key problem is how to demonstrate the “quantum advantage” in NISQ computers. In 2013, the variational quantum eigensolver (VQE) algorithm which solves Hamiltonian simulation problems was proposed\cite{vqe_2014}. In the algorithm, the quantum computer cooperates with a classical computer, and the classical computer helps to optimize the parameters in a parameterized quantum circuit. Since 2014, many algorithms, such as the quantum approximate optimization algorithm (QAOA)\cite{qaoa_2014}, quantum circuit learning (QCL)\cite{qcl_2018} and quantum variational classifier\cite{havlicek_supervised_2018}, have been proposed\cite{QGAN_lloyd, QGAN_PRA_2018, q_autoencodrs, q_BoltzmannMachine, farhi_classification_2018, schuld_quantum_2018, schuld_circuit-centric_2018, schuld_circuit-centric_2018}. Together with VQE, these algorithms attempt to demonstrate the quantum advantage in limited-size quantum circuits.

These quantum algorithms designed for the NISQ computer are often programmed with a variational quantum circuit and a classical optimizer. The variational circuit often starts with some random parameters, and then, by evaluating the cost function, the optimizer updates the parameters as feedback. Repeating this finally leads to the optimum result, which is likely to be the ground energy of a Hamiltonian, or a correct classifier.

The process is similar to the parameter updating in many machine learning algorithms. Many Machine Learning (ML) frameworks exist, such as TensorFlow and pyTorch. Inspired by these frameworks, we study the possibility of implementing a mixture of the variational quantum algorithm and a classical machine learning framework. We abstract the quantum subroutine as an operator and embed it into a simple classical machine learning framework. The framework enables the user to build quantum circuits as a common operation, for example, matrix multiplication, and allows automatic differentiation and back propagation. The hybrid framework provides the ability to train a mixture of a neural network and variational quantum circuits, which makes it practical for building a deep quantum-classical hybrid neural network, which is a possible approach to compare the performance against the classical Deep Neural Network(DNN) and to demonstrate the quantum advantage.

\section{Differentiation of a Quantum Operation} \label{Section2}
A variational quantum algorithm often involves a quantum circuit ansatz, which is based on some unknown parameters. These parameters are chosen randomly at the beginning of constructing the quantum circuit ansatz.

By repeatedly performing quantum circuits many times, the result corresponds to the expectations of observables. For example, suppose $Q0$ and $Q1$ are two qubits, then sample the probabilities of four states $|00\rangle, |01\rangle, |10\rangle, |11\rangle$ in subspace constructed by $Q0$ and $Q1$, we can compute the expectation of observable $Z_0Z_1$ with formula:
$$
\langle Z_0Z_1 \rangle=p_{00}-p_{01}-p_{10}+p_{11}
$$
where $p_{ij}$ represents the probability of the state $|ij\rangle$.

Variational quantum algorithms make use of the expectations, which are further fed into other subroutines. Finally, one algorithm run generates a loss function, and the task is to minimize the loss. It is free to choose how to minimize the loss (or optimize the problem itself). There are two types of optimizing methods: gradient-based and gradient-free.

A gradient-free optimizer only evaluates the loss and adjusts the corresponding variables. A gradient-based optimizer typically evaluates both the loss and the gradient of the loss for all variables. A typical example is the well-known gradient descent optimizer. The optimizer adjusts the variables against the direction of the gradients with a preset learning rate. The learning rate is sufficiently small to guarantee that the loss decreases. The vanilla gradient descent optimizer is effective, but not always efficient. Tuning the learning rate dynamically has resulted in other types of gradient-based optimizers, for example, AdaGrad, Adam, and RmsProp.
The gradient of the loss for all variables can be computed efficiently through back propagation in classical machine learning framework, here we show the method of computing gradient of the loss in variational quantum circuits.

First, we can efficiently compute the expectation of a certain Hamiltonian. Suppose there is a Hamiltonian $ H $, which can be written as 
$$
H=\sum_{i}{C_iH_i},
$$ 
where $ C_i $ is a real number and $ H_i $ is a tensor product of Pauli operators. For any quantum state, the expectation of H is 
$$
\langle H \rangle=\sum_{i}{C_i\langle H_i \rangle}.
$$
The expectation of $ H_i $ is easy to obtain\cite{nielsen_quantum_2002}. Then if $ H $ has polynomial Pauli tensor product operators, we can efficiently obtain the expectation of $ H $.

Next, we show how to compute the gradient of the expectation of a certain Hamiltonian. Suppose the initial state is $|0\rangle^{\otimes n}$, where $n$ is the number of qubits. The variational quantum circuit is $ U(\vec{\theta})$ and the target Hamiltonian is $ H $. Then the expectation of $H$ represented with $E$, can be written as 
$$
E=\langle 0|^{\otimes n}U(\vec{\theta})^{\dagger}HU(\vec{\theta})|0\rangle^{\otimes n}.
$$
$E$ is a function of $\vec{\theta}$, so it can be written as $E(\vec{\theta})$. We show how to obtain the gradient of $E(\vec{\theta})$.

First, we consider a special case in which $ U(\theta) $ consists of a chain of unitary transformations: 
    $ \prod_{j=1}^l U_j(\theta)$, where $\theta$ is in $\{\theta_1, \theta_2,...,\theta_m\}$ , and $U_j(\theta )=e^{-i\theta C_jP_j/2} $, where $C_j $ is a constant real number and $P_j$ is a Pauli product. Then the gradient of $E(\vec{\theta})$ can be written as 
 $$
    \frac{\partial{ E(\vec{\theta}) }}{\partial{\theta_i}}=
    \sum_{j\in{\{j|\theta_i\in U_j\}}} {C_j\frac{\langle H \rangle_{ji}^+-\langle H \rangle_{ji}^-}{2}},
    $$
    where\\
    $ \langle H \rangle_{ji}^+=Tr(HU_{l:j+1}U_j(\theta_i+\pi/2)\rho_{j-1}U_j(\theta_i+\pi /2)^\dagger U_{l:j+1}^\dagger) $\\
    $ \langle H \rangle_{ji}^-=Tr(HU_{l:j+1}U_j(\theta_i-\pi/2)\rho_{j-1}U_j(\theta_i-\pi /2)^\dagger U_{l:j+1}^\dagger). $

Here we give proof of this formula. First write the definition of $E(\vec{\theta})=\langle H \rangle =Tr(HU_{l:1}\rho U_{l:1}^\dagger )$
    $$
    \frac{\partial{U_j(\theta_i)}}{\partial{\theta_i}}=-\frac{i}{2}C_jP_jU_j(\theta_i).
    $$
    Then,
    
    \begin{align*}
    \frac{\partial{ E(\vec{\theta})}}{\partial{\theta_i}}=\left({\sum_{j\in{\{j|\theta_i\in U_j\}}} {-\frac{i}{2}C_jTr(HU_{l:j}(P_jU_{j-1:1}\rho_{in} U_{j-1:1}^\dagger}} \right. \\ 
    \left. {-U_{j-1:1}\rho_{in} U_{j-1:1}^\dagger P_j)U_{l:j}^\dagger )} \right) \\
    \end{align*}   
    $$    
    \\
    \frac{\partial{ E(\vec{\theta})}}{\partial{\theta_i}}=\sum_{j\in{\{j|\theta_i\in U_j\}}} {-\frac{i}{2}C_jTr(HU_{l:j}[P_j,U_{j-1:1}\rho_{in} U_{j-1:1}^\dagger]U_{l:j}^\dagger )}. 
    $$
    Next based on the property of the commutator for an arbitrary operator $\rho $\cite{li_hybrid_2017},  
    $$
    [\sigma_\alpha,\rho]=i\left[ R_\alpha\left(\frac{\pi}{2}\right)\rho R_\alpha\left(\frac{\pi}{2}\right)^\dagger -  R_\alpha\left(-\frac{\pi}{2}\right)\rho R_\alpha\left(-\frac{\pi}{2}\right)^\dagger\right].
    $$
    We can obtain  
    $$
    \frac{\partial{E(\vec{\theta})}}{\partial{\theta_i}}=\sum_{j\in{\{j|\theta_i\in U_j\}}} {C_j\frac{\langle H \rangle_{ji}^+-\langle H \rangle_{ji}^-}{2}}.
    $$
Furthermore, arbitrary variational quantum circuit $U$ can be represented or decomposed into  $ \prod_{j=1}^l U_j(\theta), \theta\in\{\theta_1,\theta_2,...,\theta_m\}$\cite{nielsen_quantum_2002}, so we can efficiently compute the partial derivative for a variable to the expectation any variational quantum circuit, and the circuit framework is unchanged when computing the partial derivative, we just change the parameters of the variational quantum circuit. Using this way, we obtain the gradient of the loss. 

In the specific implementation, we define a VQC object to store the information of a variational quantum circuit. One variable can be related to different gate in a VQC, so each VQC object holds a map to the inner variable and the corresponding gate, then we can compute the partial derivative for the variable based on the way described in this section. 

In next section, we define two quantum operators which support automatic differentiation using the method described in this section.

\section{Quantum Operator} \label{VQOP}

A classical machine learning framework is typically a symbolic computing system. The system allows the user to define variables and their computing rules (e.g., matrix multiplication, cross entropy or reduce sum). In the last part of classical machine learning framework, all the input flows into a loss (or reward, in a different ML context). The ML framework can then use an optimizer (e.g., gradient descent optimizer) to minimize the loss function. This is how a ML task works.

\begin{figure}[htbp]
    \includegraphics[width=0.45\textwidth]{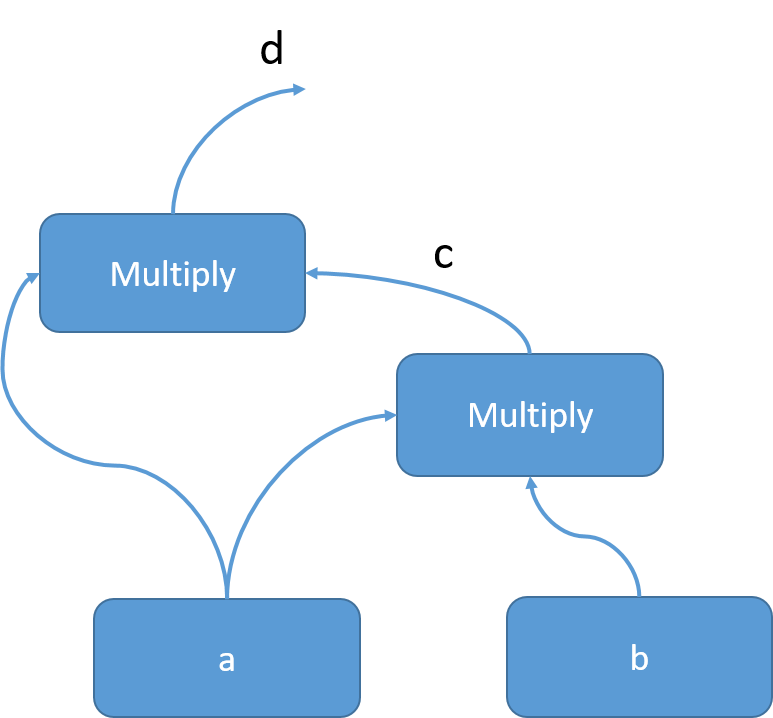}
    \caption{Example of a symbolic computing graph: $d=a\times c$ and $c=a\times b$}
    \label{SybmolicComputingGraph}
\end{figure}

In the ML framework, the computing procedures are not directly computed, but add a node to a graph that is connected to the input nodes. The graph represents a symbolic computing system. For example, $a$ and $b$ are two variables, $c=a\times b$ and $d=a\times c$ are expressed by the graph in Fig. \ref{SybmolicComputingGraph}

The gradient of the loss to the variables can be computed by a back-propagation graph. The partial derivatives of $d$ on $a$ and $b$ can be directly computed through the graph.

A quantum problem, which is the expectation of a Hamiltonian and a VQC, can also be considered as an operator in the ML framework. Suppose a variational quantum circuit is programmed as in Fig.\ref{SybmolicComputingGraph}. The quantum computer processes the circuit with all parameters fed (thus, it becomes a typical quantum circuit), and measures the final state several times to obtain the expectations in the Z-basis.

$$
E=\langle 0|C^{\dagger} HC|0 \rangle=\langle\psi|H|\psi\rangle.
$$
1.	One can choose more than one Hamiltonian, which results in an extra cost.
2.	X, Y, or other bases can be considered as a part of the VQC.

Here we define a kind of quantum operator named QOP. There are two different QOPs: $qop$ and $qop\_pmeasure$.

First, $qop$ is a QOP which takes a VQC and several Hamiltonians as input, and outputs the expectations of the input Hamiltonians. If $w$ denotes the variables of the VQC, then 
$$
y = qop(VQC(w),Hamiltonians),
$$
Second, $qop\_pmeasure$ is another QOP. It is similar to $qop$. The input of $qop\_pmeasure$ consists of a VQC, measured qubits and components. It can compute the probabilities of all projection states in the subspace constructed by measured qubits and return some of them. The components store the labels of target projection states. It is obvious that the probability of any projection state can be regarded as an expectation of a Hamiltonian, so $qop\_pmeasure$ is a special case of $qop$.
In next section, we introduce a Hybrid quantum-classical machine learning framework, VQNet, which contains $qop$ and $qop\_pmeasure$.

\section{Hybrid quantum-classical Machine Learning Framework}

By abstracting the quantum subroutine as an operator, it can be merged into the current machine learning framework.

For example, a typical method to classify the MNIST handwritten digits is to use a convolutional neural network (CNN), which can be a convolutional layer followed by fully connected layers. By replacing the fully connected layers by a quantum classifier, the conventional CNN is “quantized.” In this paper, we do not study the advantage that the quantum concept provides, but provide the possibility of constructing a quantum-classical hybrid algorithm and make use of the power of an NISQ computer.

We propose a hybrid quantum-classical machine learning framework, which is called a variational quantum network (VQNet). VQNet is built in Quantum Programming Architecture for NISQ device application(QPANDA). QPANDA is a package of quantum
programming interfaces developed by Origin Quantum. it is mainly written in C++, and has both C++ and Python APIs\cite{chen_qrunes:_2019}. In the following examples, We use the Python APIs of QPANDA, pyQPanda, as a tool to show some details of VQNet. First, we show the core components of VQNet.

\subsection{Variable}

Variable is a data type in VQNet that is used to store variables of a specific hybrid quantum-classical network. Often, the task is to optimize the variables to minimize the cost function. A variable can be a scalar, a vector or a matrix. For example,

\begin{lstlisting}[language=Python,numbers=none]
# Define a scalar variable, init value is 2.3.
v_scalar=var(2.3)   
# Define a vector variable, dimension is 4, 
# initial value is random.
# np represents numpy module in python
v_vector=var(np.random.random((4,))   
# Define a matrix variable, dimension is (4,3),
# initial value is random.
v_matrix=var(np.random.random((4,3)))       
\end{lstlisting}

The variable has a tree structure. It can have children or parents. If a variable has no children, then we call it a leaf. We can set specific value to a variable, on the other hand, we can also obtain the value of a variable if all leaves of the variable have been set value. For example:
\begin{lstlisting}[language=Python,numbers=none]
# Get the value of v_scalar.
value=v_scalar.get_value() 
# Set value of v_vector.
v_vector.set_value(np.zeros((4,1)) 
\end{lstlisting}
\subsection{Placeholder}

For the task of machine learning, not only variables, but also placeholders are required to hold a variety of input training features. Similar to TensorFlow, in VQNet, the user can define a placeholder with shape $[None, m]$, where the first shape size relates to the batch of training data, and the second shape size is considered as the shape of the training feature.  The first shap size of a placeholder is changeable, it depends on the batch of training data. For example, for a MNIST dataset that has 10,000 samples, we can define the training data and training label as follows:

\begin{lstlisting}[language=Python,numbers=none]
# Define training data placeholder.
training_data=placeholder(shape=(784,1)) 
# Define training label placeholder.
training_label=placeholder(shape=(10,1)) 
# Feed placeholder, assume data is a
# training dataset which has 10000 samples,
# and label is a training label
# which has 10000 samples.
# Feed data to training_data 
# and label to training_label.
training_data.feed(data)
training_label.feed(label)

\end{lstlisting}
\subsection{Operator}
VQNet contains many types of operators. These operators operate on variables or placeholders. 
Some operators are classical operators, such as plus, minus, multiply, divide, exponent, log, and dot. VQNet contains all common operators from classical machine learning. VQNet also has quantum operators, which are $qop$ and $qop\_pmeasure$. $qop$ and $qop\_pmeasure$ have been described in Section \ref{VQOP}. In addition, $qop$ and $qop\_pmeasure$ are related to quantum computer chips, so they need a quantum environment. To achieve this, we need to add extra parameters when we use these two operators. In general, we add reference of quantum machine and allocated qubits to provide a quantum environment. We provide some examples as follows:

\begin{lstlisting}[language=Python,numbers=none]
# Define a virtual quantum machine, there are three 
# types of virtual quantum machine in pyQPanda:
# CPU_SINGLE_THREAD,CPU and GPU.
machine=init_quantum_machine(
  QuantumMachine_type.CPU_SINGLE_THREAD)
# Allocate 5 qubits from machine.
qlist=machine.qAlloc_many(5)
# Define a Hamiltonian H=Z0+2.4*X1Y2, 
# X,Y,Z represents Pauli X,Y,Z matrix
# respectively and the number after X/Y/Z 
# represents qubits number.
Hamiltonian=PauliOperator({'Z0':1,'X1 Y2':2.4})
# Define a QOP. 
qop_loss=qop(vqc,Hamiltonian,machine,qlists)
# Define a qop_pmeasure.
qpm_loss=qop_pmeasure(vqc,
                       [1,2,4],
                       machine,
                       [qlist[0],qlist[2],qlist[4]])
\end{lstlisting}

\subsection{Architecture of VQNet}

Variable, placeholder and operator form a symbolic computing system in VQNet. This system contains classical components and quantum components. User can construct hybrid quantum-classical machine learning network with this system. The system is a tree architecture, the root of the tree is a variable, it usually represents the cost function of a certain problem. The leaves of the tree are variables or placeholders. The steps to construct quantum-classical hybrid machine learning network with VQNet are as follows: First, determine the cost function of the target problem. Second, construct the VQC of the target problem and use the VQC and input Hamiltonians to construct QOP. Then using variables, placeholders and operators to construct the variable which represents the cost function. 

VQNet supports forward and back propagation. Before we execute a quantum-classical hybrid machine learning network, we need to feed all placeholders of the network and evaluate the network, which means feed all variables of the network, in particular, after the variables of VQC being fed, the VQC becomes a typical quantum circuit. Then we can execute forward propagation and get the root value of the network, in other words, get the value of the cost function.

When performing back propagation in VQNet, we define an $Expression$ with the root variable of the network. $Expression$ is a package for variable, it is an object which holds the information of the variable and supports forward and back propagation, we execute back propagation on the $Expression$ and obtain the gradient of the cost function. 

Finally, after getting the gradient of the cost function, we choose an optimizer to use the gradient to optimize the variables of the network. By this way, we obtain the optimum value of the variables and solve the given problem. Here we give an example:

\begin{lstlisting}[language=Python,numbers=none]
def network_sample(    
    data,        # training data    
    H,           # feature Hamiltonians    
    theta,       # Variable    
    machine,     # Reference of the quantum machine    
    qubits,      # Reference of the qubits 
):
    # Define an empty VQC  
    vqc=VariationalQuantumCircuit()    
    # Generate VQC with specific     
    # user defined VQC generation function    
    vqc_generation(vqc,data,theta,qubits)    
    # Construct QOP.    
    loss=qop(vqc,H,machine,qlists)    
    # Construct Expression.
    cost_function=expression(loss)    
    # Momentum optimizer.
    learning_rate=0.02    
    momentum=0.9     
    opt=MomentumOptimizer(cost_function,                            
                                learning_rate,
                                momentum)    
    # Define maximal iteration times.    
    max_iterations=200    
    # Run optimizer     
    opt.run(max_iterations)    
    # Get cost function value after optimization.    
    optimized_cost_function=opt.get_value()    
\end{lstlisting}

In next section, we give some examples of specific problems and show the ability of VQNet to construct quantum-classical hybrid machine learning network.

\section{Example}\label{section:example}
\subsection{QAOA algorithm for the MAX-CUT problem}
The QAOA is a well-known quantum-classical hybrid algorithm\cite{qaoa_2014}. In an algorithm run, the QAOA algorithm has the following circuit ansatz.

The $\gamma$ and $\beta$ are the unknown variables. $H_p$ and $H_d$ are the problem Hamiltonian and the driver Hamiltonian respectively, which is an analog to the adiabatic algorithm. 
For an n-object MAX-CUT problem, $n$ qubits are needed to encode the result, where the measurement results (a binary string) denote a cut configuration of the problem\cite{qaoa_rigetti_2017, qaoa1_2018, qaoa2_2018, qaoa3_2018}.
The quantum circuit is built in pyQPanda as follows:
\begin{lstlisting}[language=Python,numbers=none]
def QAOA(
    circuit,      # Reference of the circuit
    Hp,           # Problem Hamiltonian
    Hd,           # Driver Hamiltonian
    gamma,        # Variable
    beta,         # Variable
    n_step,       # Step of QAOA
    qubits        # Reference of the qubits
):
    # First apply Hadamard to all qubits
    circuit.insert(H(qubits))
    # Apply evolution on Hp and Hd with
    # the corresponding variable 
    # (gamma and beta) in each step.
    for i in range(n_step):		
        evolution(circuit, Hp, gamma[i], qubits)
        evolution(circuit, Hd, beta[i], qubits) 
\end{lstlisting}

\begin{figure}[htbp]
    \includegraphics[width=0.45\textwidth]{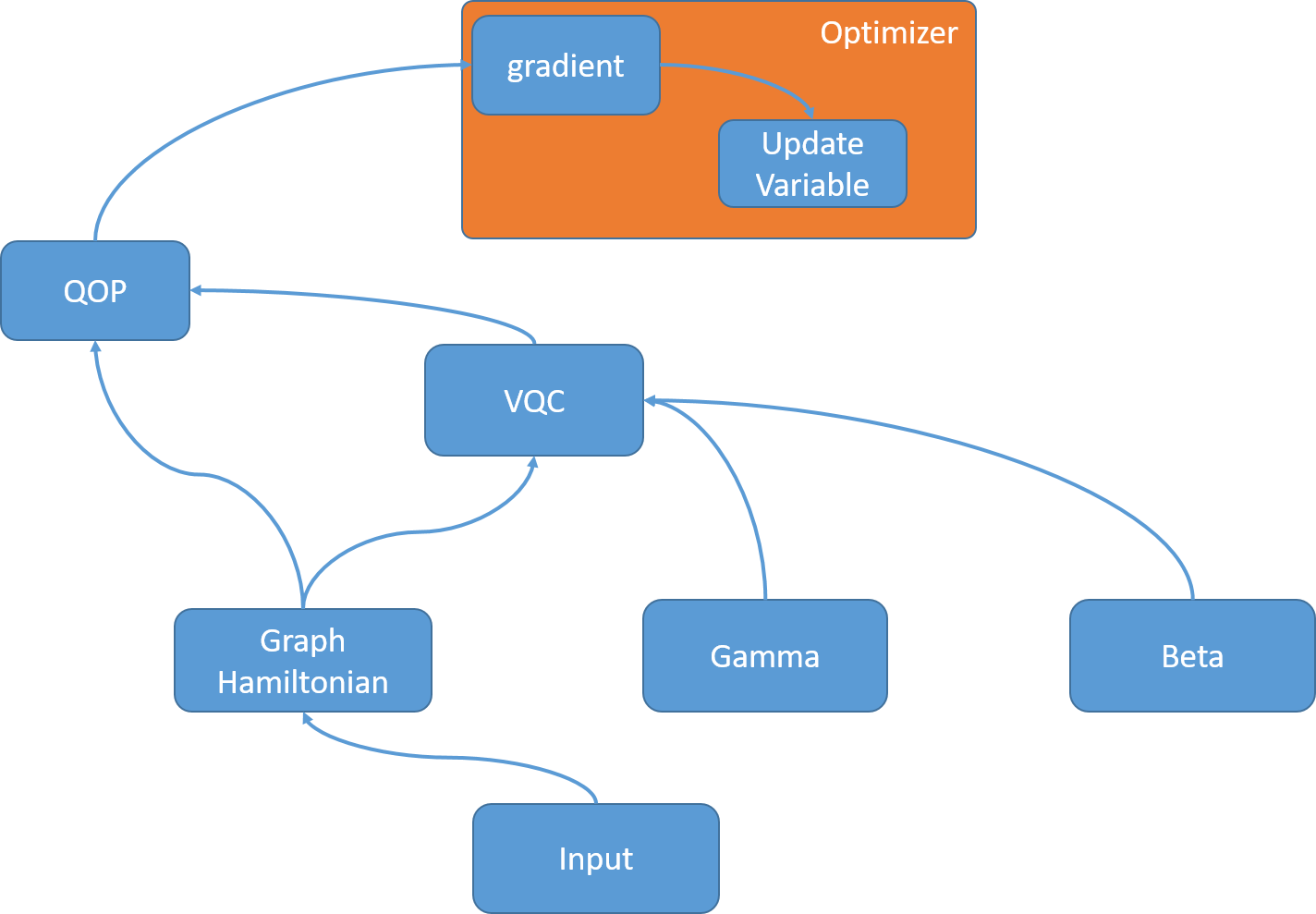}
    \caption{QAOA flow chart in VQNet}
    \label{QAOAFlowChart}
\end{figure}

We can efficiently realize QAOA for the MAX-CUT problem in VQNet. The flow chart of QAOA in VQNet is shown in Fig.\ref{QAOAFlowChart}. 

First, we input the graph information of the MAX-CUT problem and construct the corresponding problem Hamiltonian and driver Hamiltonian. Then, using the Hamiltonian and variable $Gamma$, $Beta$ , we construct the VQC of QAOA. The input parameters of the QOP are the problem Hamiltonian and the VQC, and the output of the QOP is the expectation of the problem Hamiltonian. In this problem, the cost function is the expectation of the problem Hamiltonian, so we need to minimize the output of the QOP. We can obtain the gradient of the cost function through back propagation and use the optimizer to update $Gamma$ and $Beta$ of the VQC. The QAOA flow chart is built in pyQPanda as follows:

\begin{lstlisting}[language=Python,numbers=none]
def QAOA_flowchart(
    Hp,          # Problem Hamiltonian
    Hd,          # Driver Hamiltonian
    gamma,       # Variable
    beta,        # Variable
    n_step,      # Step of QAOA
    machine,     # Reference of the quantum machine
    qubits,      # Reference of the qubits
):
    # Define an empty VQC.
    vqc=VariationalQuantumCircuit()
    # Construct quantum circuit of QAOA.
    QAOA(vqc,Hd,Hp,gamma,beta,n_step,qubits)
    # Construct QOP.
    loss=qop(vqc,Hp,machine,qlists)
    # Transform loss to an Expression.
    cost_function=expression(loss)
    # Momentum optimizer
    learning_rate=0.02
    momentum=0.9 
    opt=MomentumOptimizer(cost_function,
                            learning_rate,
                            momentum)
    # Define maximal iteration times.
    max_iterations=200
    # Run QAOA 
    opt.run(max_iterations)
    # Get cost function value after optimization.
    optimized_cost_function=opt.get_value()  
\end{lstlisting}

\begin{figure}[htbp]
    \subfigure[Max-Cut Graph]{\includegraphics[width=0.30\textwidth]{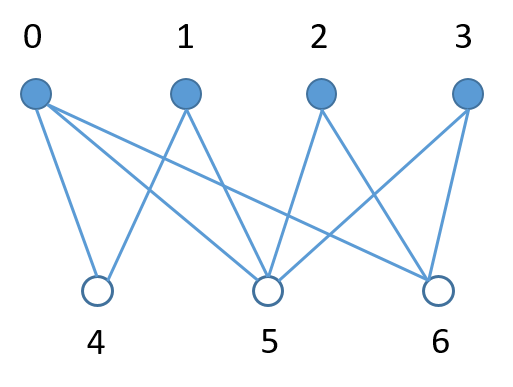}}
    \subfigure[Weight of Graph Edge]{\includegraphics[width=0.30\textwidth]{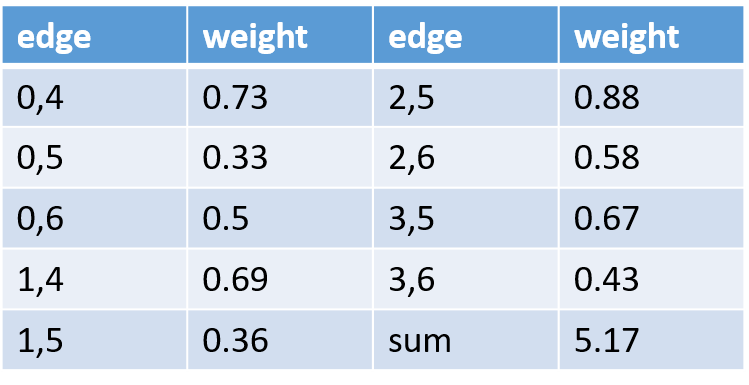}}
   
    \caption{MAX-CUT graph. The figure (a) is the shape of the MAX-CUT graph and the figure (b) is the edge weight of the graph. The MAX-CUT value is 5.17.}
    \label{MaxCutGraph}
\end{figure}

Here we use VQNet to solve a specific MAX-CUT problem, the graph of the problem is shown in Fig.\ref{MaxCutGraph}, its MAX-CUT value is 5.17, and the optimum division plan is vertex 0,1,2,3 are in one group and the other vertices are in the other group.

\begin{figure}[htbp]
    \subfigure[]{\includegraphics[width=0.22\textwidth]{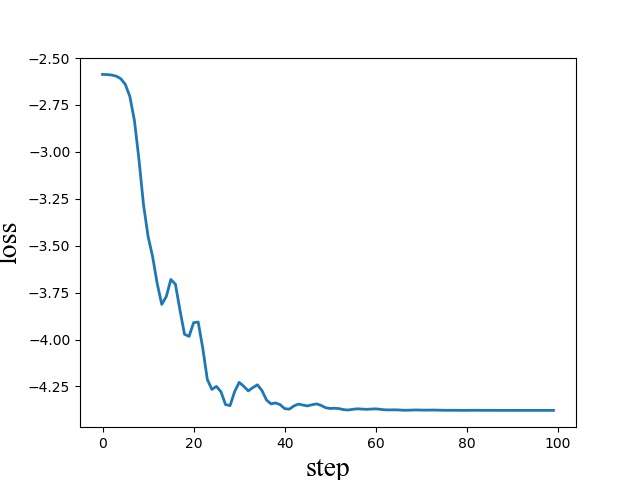}}
    \subfigure[]{\includegraphics[width=0.22\textwidth]{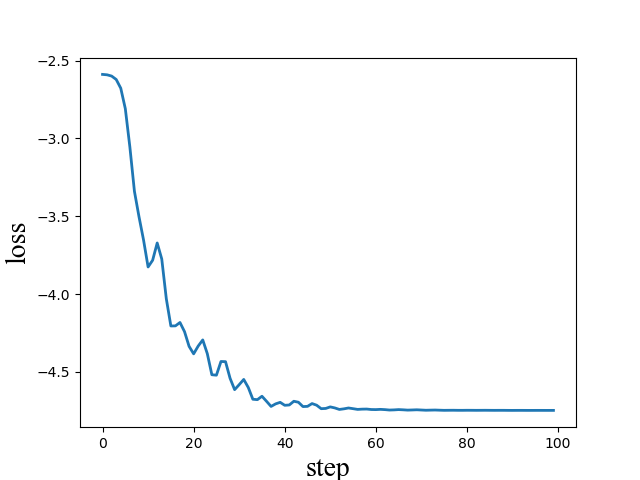}}
    \subfigure[]{\includegraphics[width=0.22\textwidth]{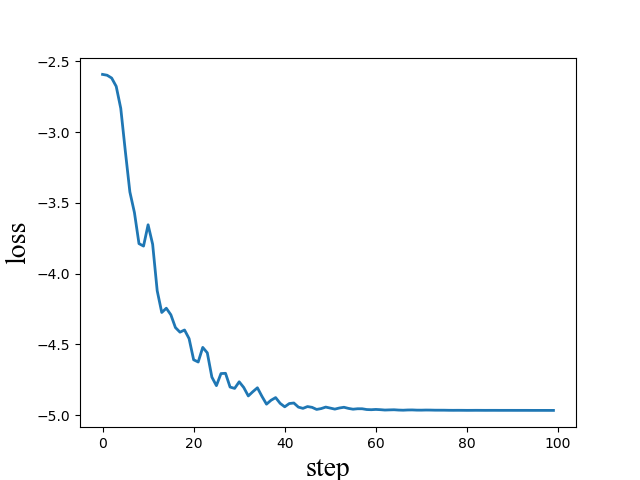}}
    \subfigure[]{\includegraphics[width=0.22\textwidth]{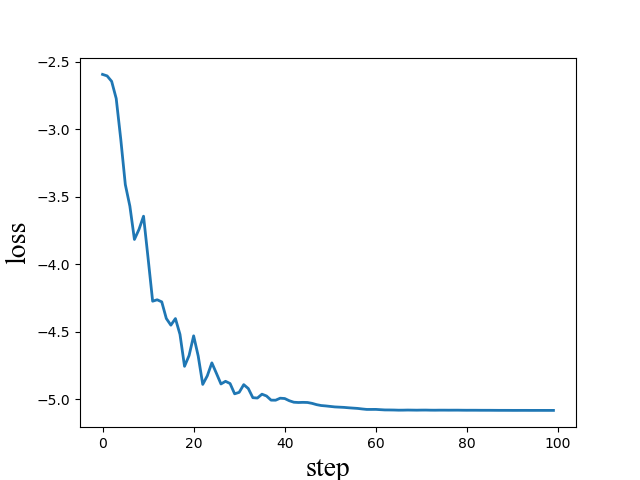}}
   
    \caption{Relation between the cost function and training step, in figure (a), (b), (c), (d), the QAOA steps are 2,3,4,5, respectively. The ratios of the target value and trained value is 0.8468, 0.9183, 0.9605, 0.9830, respectively.}
    \label{QAOA Optimization}
\end{figure}

We use the momentum gradient descent optimizer\cite{deeplearning-Goodfellow-et-al-2016}. The QAOA step is set to 2, 3, 4, 5 respectively. The optimization process is shown in Fig.\ref{QAOA Optimization}.

\subsection{Variational Quantum Eigensolver}
\begin{figure}[htbp]
    \includegraphics[width=0.45\textwidth]{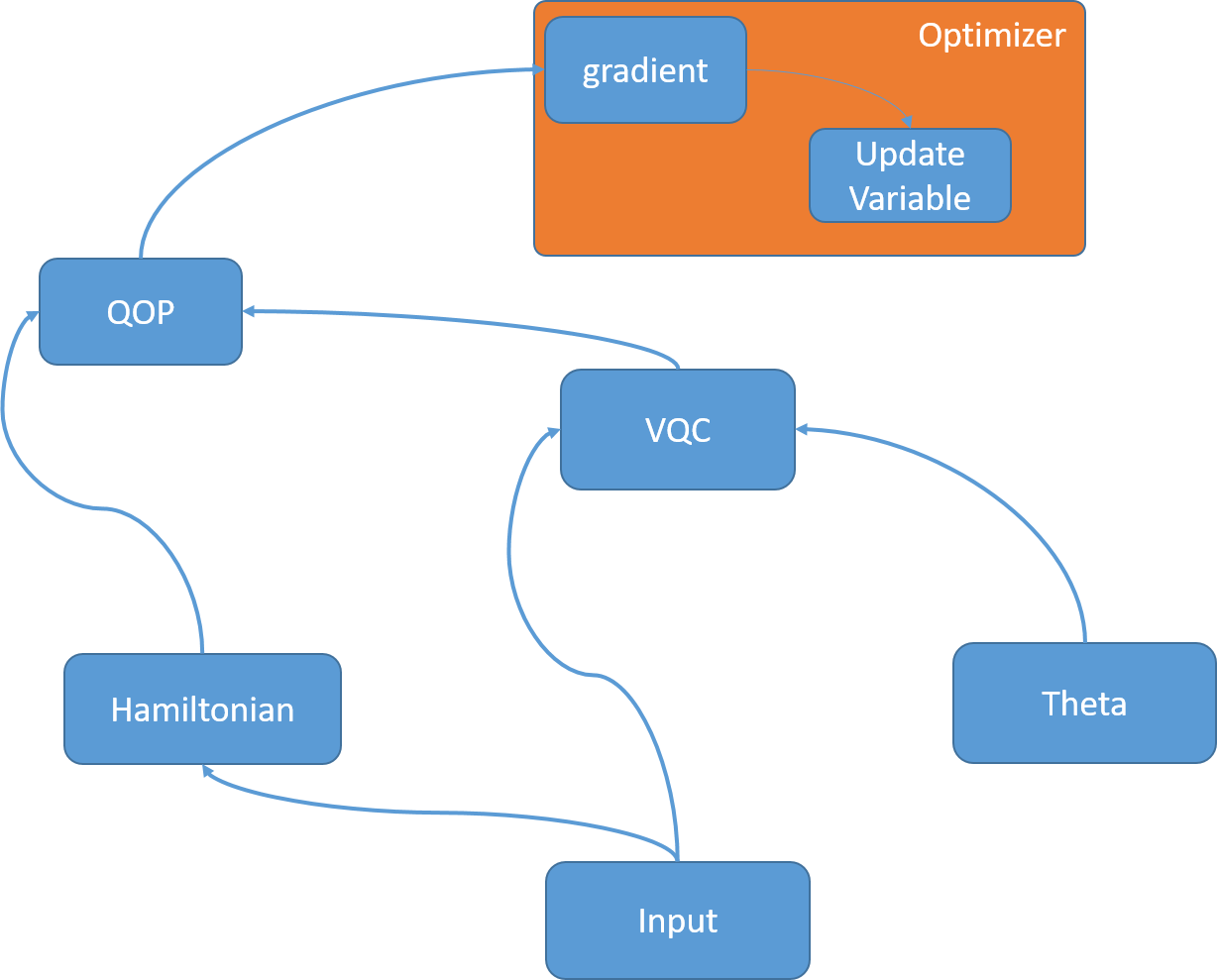}
    \caption{VQE algorithm flow chart in VQNet}
    \label{VqeAlgorithm}
\end{figure}

The Variational Quantum Eigensolver(VQE) algorithm is another prospective NISQ quantum algorithm. It can be used in quantum chemistry. Many work related to the VQE algorithm has been proposed in recent years\cite{vqe_2014,omalley_scalable_2016, kandala_hardware-efficient_2017}. Here we present a VQE algorithm example with VQNet.
The VQE algorithm flow chart in VQNet is shown in Fig.\ref{VqeAlgorithm}.
In the VQE algorithm, we use the VQC architecture proposed in \cite{kandala_hardware-efficient_2017}, $Theta$ is the variable of the VQC. The user inputs the molecule Hamiltonian, then the molecule Hamiltonian and the VQC construct QOP. The target of the VQE algorithm is to make the quantum state evolve to the ground state of the molecule Hamiltonian, so we only need to minimize the output of the QOP, that is, the expectation of the molecule Hamiltonian. 
\begin{lstlisting}[language=Python,numbers=none]
def VQE_algorithm(
    Hamiltonian, # Feature Hamiltonian
    theta,       # Variable
    machine,     
    qubits,     
): 
    vqc=VariationalQuantumCircuit()
    # Construct quantum circuit of VQE algorithm,
    # the architecture of the quantum circuit is 
    # packaged in function vqe_circuit
    vqe_circuit(vqc,theta)
    # Construct QOP.
    loss=qop(vqc,Hamiltonian,machine,qlists)
    # Transform loss to an Expression.
    cost_function=expression(loss) 
    learning_rate=0.02
    momentum=0.9 
    # Momentum optimizer
    opt=MomentumOptimizer(cost_function,
                            learning_rate,
                            momentum)
    max_iterations=200
    # Run VQE algorithm 
    opt.run(max_iterations)
    optimized_cost_function=opt.get_value()  
\end{lstlisting}

\begin{figure}[htbp]
    \subfigure[]{\includegraphics[width=0.40\textwidth]{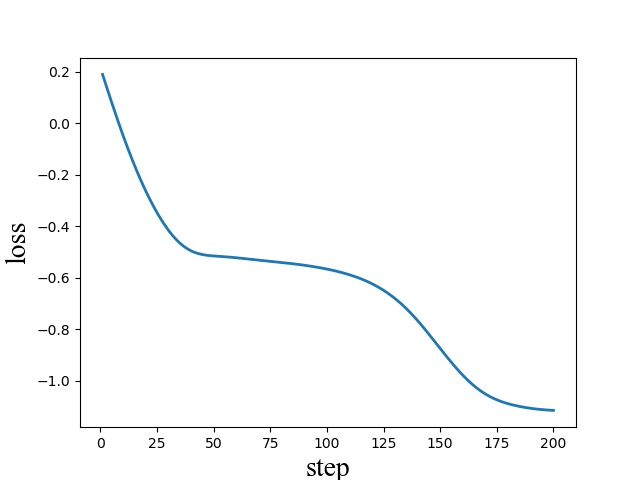}}
    \subfigure[]{\includegraphics[width=0.40\textwidth]{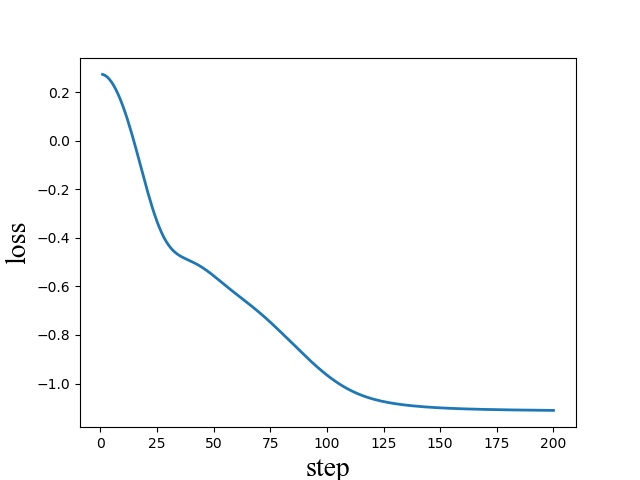}}
    \caption{Figure (a): Energy optimization process in the Adam optimizer. The figure describes the relation between the optimization step and simulated potential energy. The distance between two hydrogen atoms is $r=0.7 \AA$. Figure (b): Energy optimization process in the momentum optimizer. The figure describes the relation between the optimization step and simulated potential energy. The distance between two hydrogen atoms is $r=0.7 \AA$.}
    \label{VqeOptimizationProcess}
\end{figure}

We present a simple case, $H_2$, in which the distance between two hydrogen atoms is $r=0.7\AA$. The optimization process is shown in Figs.\ref{VqeOptimizationProcess}.

\begin{figure}[htbp]
    \includegraphics[width=0.40\textwidth]{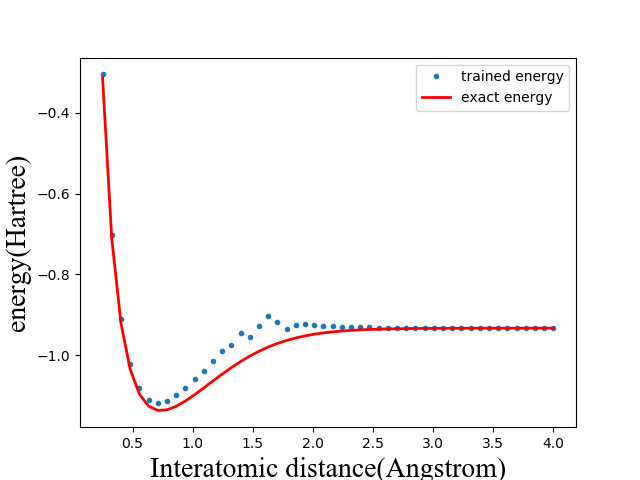}
    \caption{Potential energy surfaces: potential energy as a function of interatomic distance. The red line indicates the exact energy and the blue dotted lines indicate the simulated energy using the VQE algorithm.}
    \label{PotentialEnergySurfaces}
\end{figure}

We scan the distance in the range $0.25\AA$ to $4.0\AA$ and take 50 points evenly in this range. Then we obtain the relation between distance and energy, as shown in Fig.\ref{PotentialEnergySurfaces}.

\subsection{Quantum classifier}
\begin{figure}[htbp]
    \includegraphics[width=0.45\textwidth]{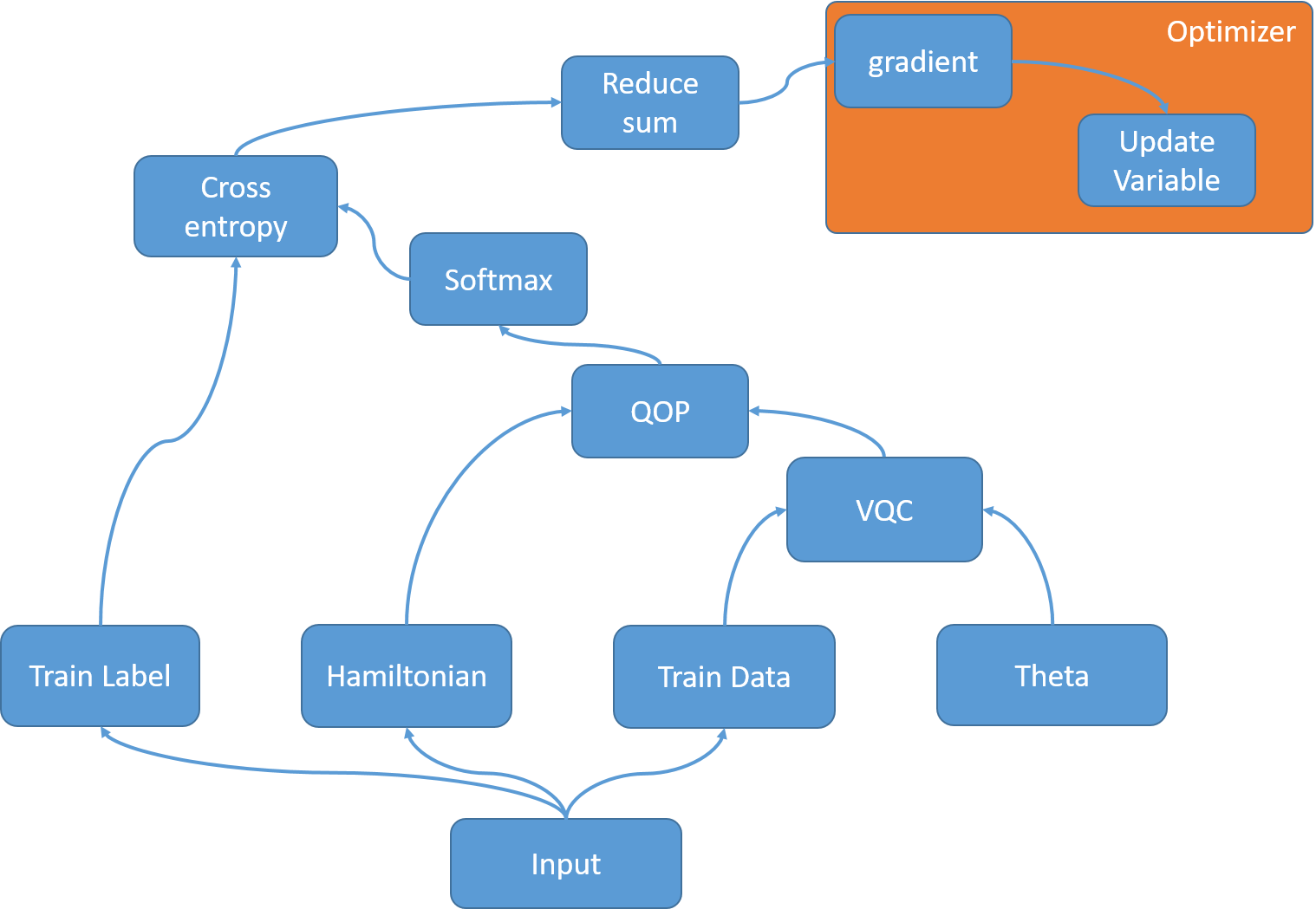}
    \caption{Quantum classifier flow chart in VQNet}
    \label{QuantumClassifier}
\end{figure}

Classification problems are common in machine learning. Many models exist for classification in machine learning, such as SVM, LR, and CNN. In recent years, some quantum algorithms for classification problems have been proposed\cite{havlicek_supervised_2018, farhi_classification_2018, schuld_quantum_2018, schuld_circuit-centric_2018, grant_hierarchical_2018}. Most of them are based on specific VQC. We use VQNet to construct a quantum classifier. The flow chart of the quantum classifier is shown in Fig.\ref{QuantumClassifier}.

The quantum classifier consists of some operations. First, we feed the input data into the classifier. The input data includes training data, training label and input Hamiltonian. The training data and variable $Theta$ are used to construct a VQC. Then the VQC and input Hamiltonian flow to the QOP, and the output of QOP is the expectation of the input Hamiltonian. Next, execute Softmax Operation on the output of the QOP. Then compute the cross entropy between the training label and outputs of the Softmax operation. Finally, the Reduce Sum operation calculates the sum of the cross entropy and the output is regarded as the cost function of the calssifier\cite{deeplearning-Goodfellow-et-al-2016}. 

We can minimize the loss function with different optimizers. In this case, we use gradient-based optimizer, so we need to obtain the gradient of the cost function. We execute back propagation to obtain the gradient and use the gradient to optimize $Theta$. Finally, we can obtain a trained quantum classifier. The corresponding pyQPanda code is shown as follows:

\begin{lstlisting}[language=Python,numbers=none]
def quantum_classifier(
    Hamiltonian,     # Feature Hamiltonian
    train_data,      # Training data
    train_label,     # Training label
    theta,           # Variable
    machine,        
    qubits,        
):
    vqc=VariationalQuantumCircuit()
    # Construct the VQC of quantum classifier
    # with function quantum_classifier_circuit.
    quantum_classifier_circuit(vqc,train_data,theta)
    # Construct QOP.
    expectation=qop(vqc,Hamiltonian,machine,qlists)
    # Operate softmax operation on expectation
    loss_softmax=soft_max(expectation)
    # Compute cross entropy of 
    # training label and loss_softmax
    entropy=cross_entropy(train_label,loss_softmax)
    # Sum the cross entropy of all train_data sample
    # with Reduce_sum operation
    loss=reduce_sum(entropy)
    # Transform loss to an Expression
    cost_function=expression(loss)
    # Momentum optimizer.
    learning_rate=0.02
    momentum=0.9 
    opt=MomentumOptimizer(cost_function,
                            learning_rate,
                            momentum)
    max_iterations=200
    # Run quantum classifier 
    opt.run(max_iterations)
    optimized_cost_function=opt.get_value()
    
\end{lstlisting}

Now we provide a specific example.

We choose breast cancer datasets as the classification problem. This is a two-group classification problem with 30 features. For simplicity, we only consider the last 10 features of the training data.

The structure of the VQC in this quantum classifier is proposed in \cite{havlicek_supervised_2018}. We make some changes to the VQC based on our task. In our example, the VQC is divided into three parts: initial state circuit, parameterized circuit and parity check circuit. The initial state circuit is used to encode the training data into the quantum circuit. The parameterized circuit has a series of parameters, and the quantum classifier completes the classification task by optimizing these parameters. Finally, because breast cancer is a two-group classification problem, we obtain the classification result using the parity check result. The specific method is to add a parity check circuit at the end of the circuit and measure the target qubit of the parity check circuit. Then we obtain the classification result based on  the $z$-axis expectation $E_z$ of the target qubit. $E_z>0$ and $E_z<0$ represent two classification results.

\begin{figure}[htbp]
    \subfigure[]{\includegraphics[width=0.22\textwidth]{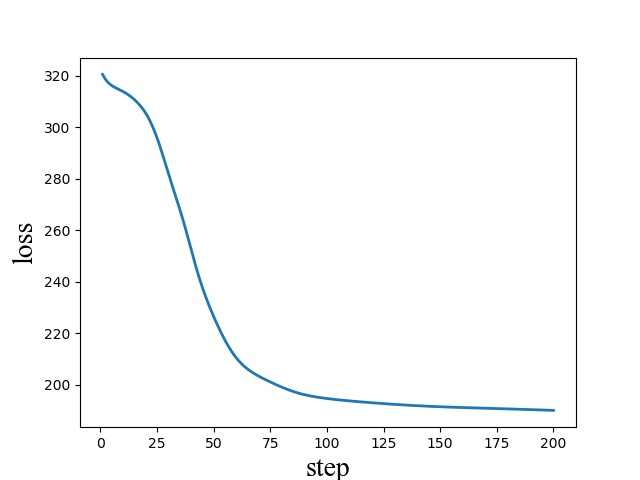}}
    \subfigure[]{\includegraphics[width=0.22\textwidth]{adam_softmax_loss.jpg}}
    \subfigure[]{\includegraphics[width=0.22\textwidth]{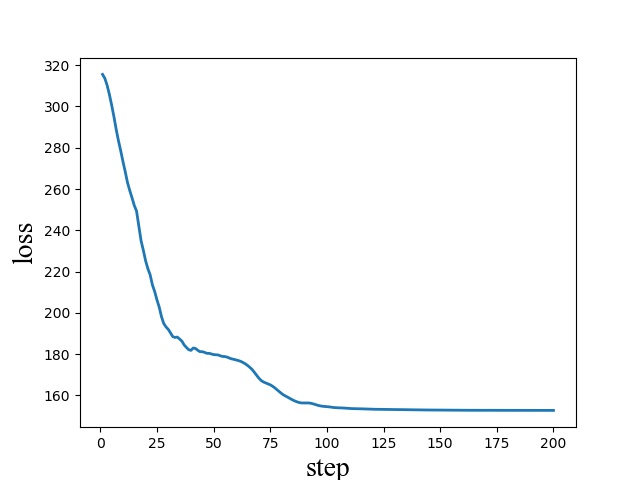}}
    \subfigure[]{\includegraphics[width=0.22\textwidth]{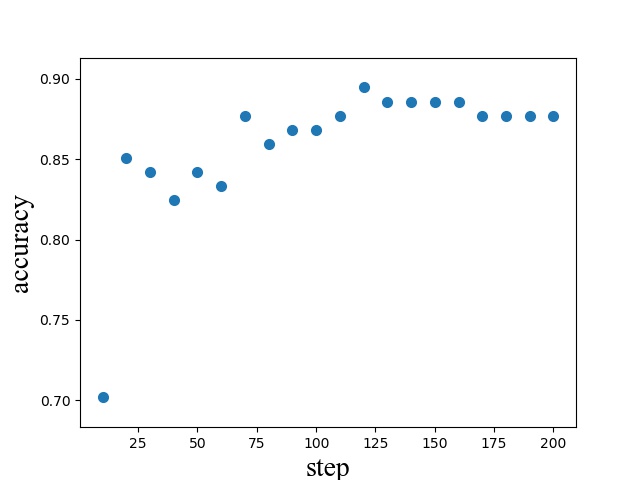}}
    \caption{Figure (a): Adam optimizer: values of the cost functions as a function of the training step; Figure (b): Adam optimizer: test accuracy as a function of the training step. Figure (c): Momentum optimizer: values of the cost functions as a function of the training step; Figure (d): Momentum optimizer: test accuracy as a function of the training step. }
    \label{QuantumClassifierOptimizationProcess}
\end{figure}

We use two optimizers to optimize the variables of the quantum classifier: momentum gradient descent optimizer and Adam optimizer. The cost function and accuracy vary with the number of optimized steps, as shown in Figs.\ref{QuantumClassifierOptimizationProcess}.

\subsection{Quantum Circuit Learning}
\begin{figure}[htbp]
    \includegraphics[width=0.45\textwidth]{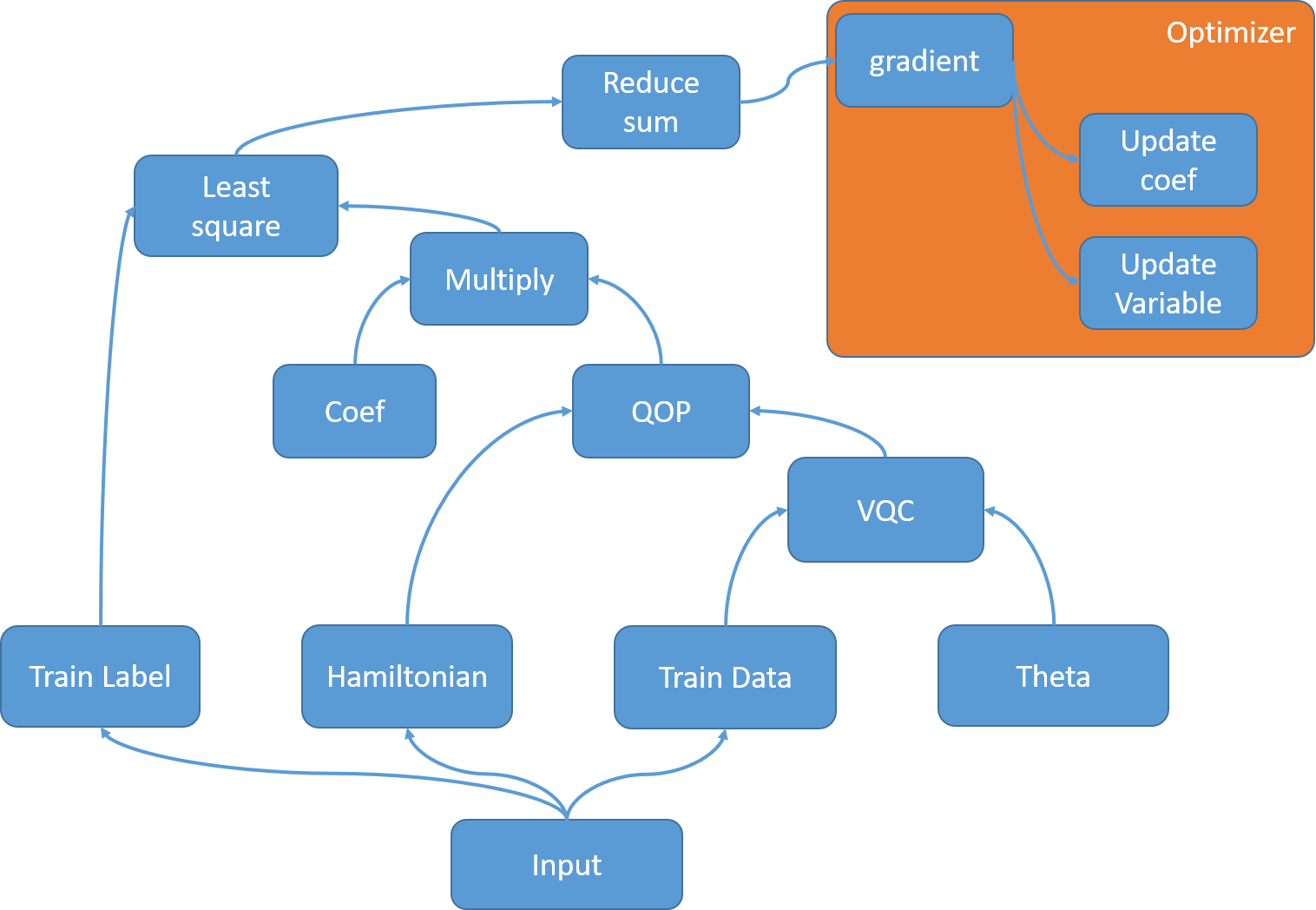}
    \caption{Quantum circuit learning flow chart in VQNet}
    \label{QuantumCircuitLearning}
\end{figure}

\cite{qcl_2018} proposed a method that variational quantum circuit can be used to learn some nonlinear functions. We construct their architecture with VQNet. The corresponding flow chart is shown in Fig.\ref{QuantumCircuitLearning}.

\begin{figure}[!h]
    \subfigure[]{\includegraphics[width=0.22\textwidth]{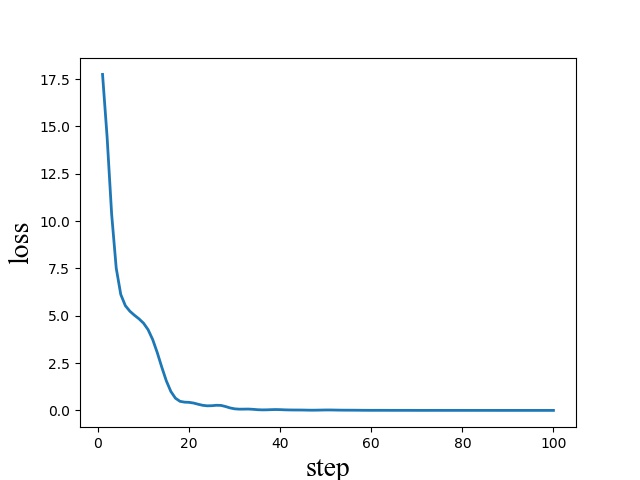}}
    \subfigure[]{\includegraphics[width=0.22\textwidth]{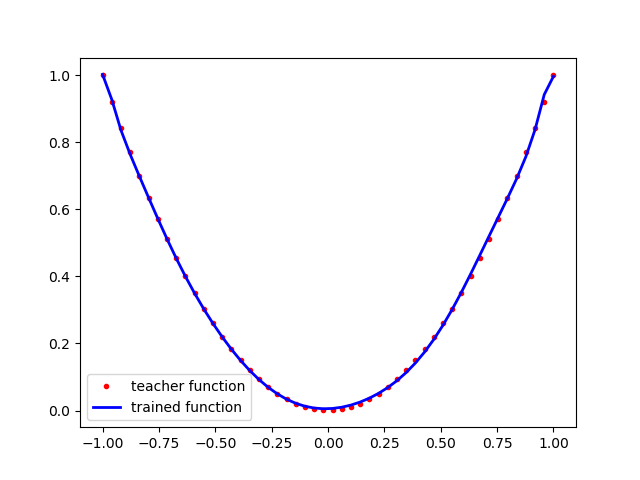}}
    \subfigure[]{\includegraphics[width=0.22\textwidth]{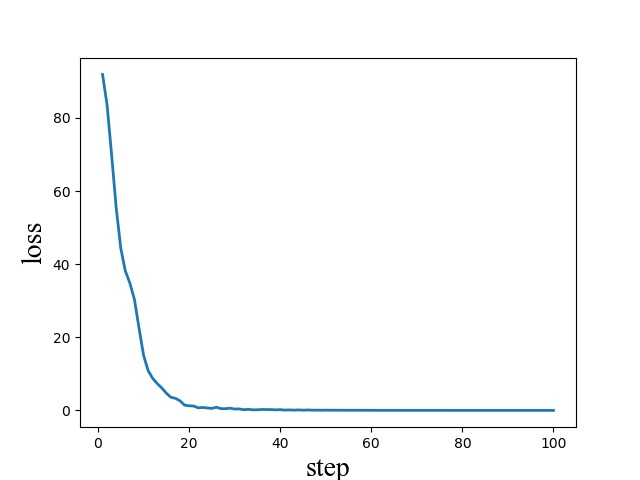}}
    \subfigure[]{\includegraphics[width=0.22\textwidth]{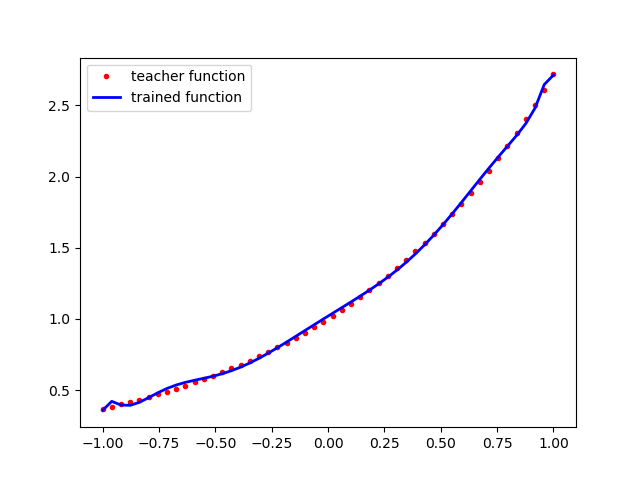}}
    \subfigure[]{\includegraphics[width=0.22\textwidth]{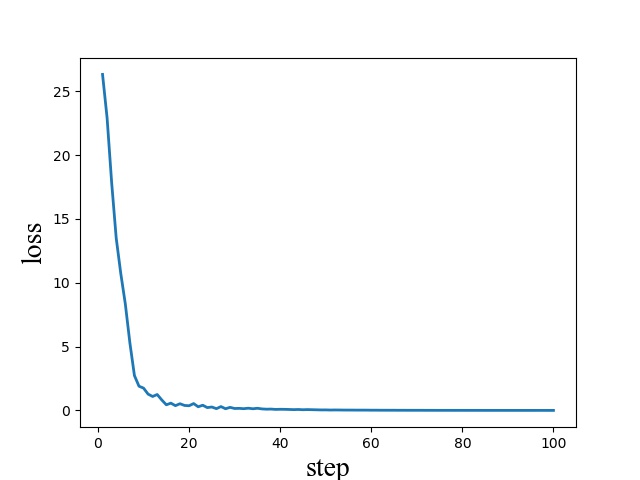}}
    \subfigure[]{\includegraphics[width=0.22\textwidth]{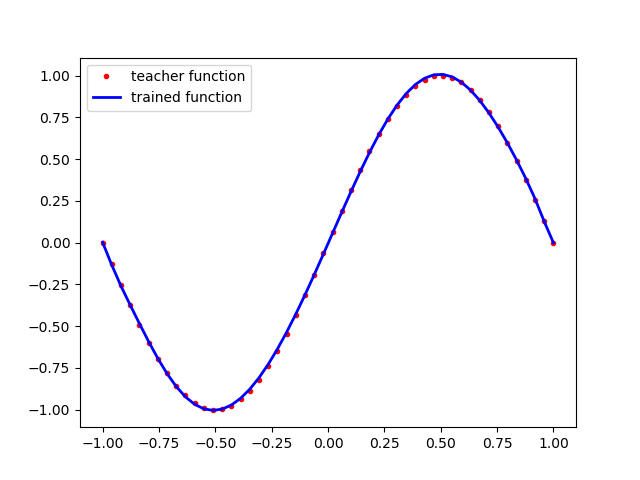}}
    \subfigure[]{\includegraphics[width=0.22\textwidth]{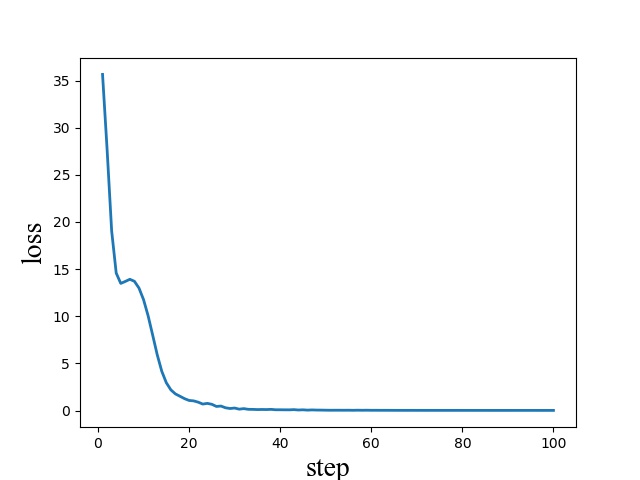}}
    \subfigure[]{\includegraphics[width=0.22\textwidth]{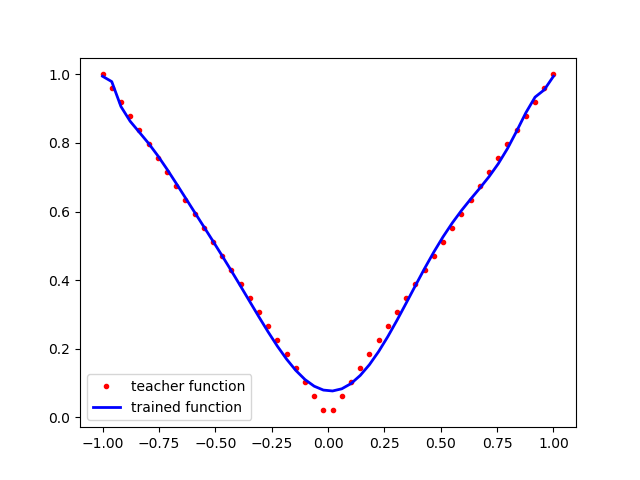}}
    \caption{cost function as a function of step and result of trained function. Figure(a),(b): training function is $y=x^2$, Figure (a) shows cost function as a function of step, Figure (b) shows result of the trained function(blue line) and teacher function(red dot line). Figure(c),(d): training function is $y=e^x$, Figure (c) shows cost function as a function of step, Figure (d) shows result of the trained function(blue line) and teacher function(red dot line). Figure(e),(f): training function is $y=sin(\pi x)$, Figure (e) shows cost function as a function of step, Figure (f) shows result of the trained function(blue line) and teacher function(red dot line). Figure(g),(h): training function is $y=|x|$, Figure (g) shows cost function as a function of step, Figure (h) shows result of the trained function(blue line) and teacher function(red dot line).}
    \label{QCLTest}
\end{figure}
The structure of the quantum circuit learning machine is similar to the classifier mentioned in previous example. Suppose the target function is $y=f(x)$, then the training data is a series of $x$ and the training label is $f(x)$. First, we use the input Hamiltonian, training data and variable $Theta$ to construct QOP. The trained function is $y_{trained}=Coef\times E$, where $E$ is the output of QOP and $Coef$ is a scalar variable. Then cost function $cost=\sum_{i}{|y_i(correct)-y_i(trained)|^2}$, and we use momentum optimizer to optimize $Coef$ and $Theta$. The corresponding realization of quantum circuit learning is as follows:

\begin{lstlisting}[language=Python,numbers=none]
def quantum_circuit_learning(
    Hamiltonian,     # Feature Hamiltonian
    train_data,      # Training data
    train_label,     # Training label
    theta,           # Variable
    coef,            # Variable
    machine,       
    qubits,       
):
    vqc=VariationalQuantumCircuit()
    # Construct the VQC of quantum circuit learning
    learning_circuit(vqc,train_data,theta)
    # Construct QOP.
    expectation=qop(vqc,Hamiltonian,machine,qlists)
    # Trained function: f(x)=coef*expectation
    train_value=coef*expectation
    # Compute square of (train_label-train_value)
    least_square_value=least_square(train_label,
                                        train_value)
    
    # Sum the least_square_value of all training 
    # data samples with Reduce_sum operation.
    loss=reduce_sum(least_square_value)
    # Transform loss into an Expression.
    cost_function=expression(loss)
    # Momentum optimizer.
    learning_rate=0.02
    momentum=0.9 
    opt=MomentumOptimizer(cost_function,
                            learning_rate,
                            momentum)
    max_iterations=200
    # Run the quantum circuit learning algorithm 
    opt.run(max_iterations)
    optimized_cost_function=opt.get_value()
    
\end{lstlisting}

In this case, we use the momentum gradient descent optimizer. We tested four types of functions: $x^2,e^x,sin(x),|x|$. The cost function and trained function are shown in Figs.\ref{QCLTest}

\section{Conclusion}

We have presented a quantum-classical hybrid machine learning architecture VQNet. VQNet efficiently connects machine learning and quantum algorithms. It provides a platform for developing and testing quantum machine learning algorithms. In conclusion, VQNet has the following features: First, VQNet contains two QOPs: $qop$ and $qop\_pmeasure$. Second, $qop$, $qop\_pmeasure$ operations and all the classical operations support forward and back propagation, so VQNet supports forward and back propagation. As a result, we can use all types of gradient-based optimizers in VQNet. Finally, from examples described in Section \ref{section:example}, we found that VQNet has the ability to construct all types of quantum machine learning algorithms.

We used VQNet to construct four quantum machine learning algorithms: QAOA, VQE algorithm, quantum classifier and quantum circuit learning. We found that VQNet worked well in constructing these quantum machine learning algorithms. 

In the future, we will extend the ability of VQNet. First, to simulate the performance ability of real quantum computer chips in VQNet, we will add noisy simulation in VQNet. Second, at present, VQNet is a bare architecture; we have completed the core components of VQNet. We will package more user-friendly features in VQNet, and include different optimizers, different cost functions and some commonly used components in the quantum machine learning algorithms. Sometimes we execute tasks with a great deal of computation, so to improve the execution speed, we need to develop VQNet on a distributed computing platform, such as GPU, TPU, or MPI. To summarize, our aim is to generate a common architecture in a quantum-classical machine learning region, similar to TensorFlow, Caffe, CNTK, and MXNet in classical machine learning.

\bibliography{ref}

\end{document}